\documentstyle[12pt]{article}
\begin{document}
\title {HIGHER-GRADE HYBRID MODEL: ENHANCEMENT OF SUPERCONDUCTIVITY}
\author {by\\MA{\L}GORZATA SZTYREN
\thanks{Department of Mathematics and Information Science,
Warsaw University of Technology, Pl. Politechniki 1, PL-00-661 Warsaw
E--mail: emes@mech.pw.edu.pl}}
\date \today
\maketitle

\input{epsf.tex}
\edef\undtranscode{\the\catcode`\_} \catcode`\_11
\newbox\box_tmp 
\newdimen\dim_tmp 
\def\jump_setbox{\aftergroup\after_setbox}
%
%
\def\resize
    #1
    #2
    #3
    #4
    {%
    \dim_r#2\relax \dim_x#3\relax \dim_t#4\relax
    \dim_tmp=\dim_r \divide\dim_tmp\dim_t
    \dim_y=\dim_x \multiply\dim_y\dim_tmp
    \multiply\dim_tmp\dim_t \advance\dim_r-\dim_tmp
    \dim_tmp=\dim_x
    \loop \advance\dim_r\dim_r \divide\dim_tmp 2
    \ifnum\dim_tmp>0
      \ifnum\dim_r<\dim_t\else
        \advance\dim_r-\dim_t \advance\dim_y\dim_tmp \fi
    \repeat
    #1\dim_y\relax
    }
\newdimen\dim_x    
\newdimen\dim_y    
\newdimen\dim_t    
\newdimen\dim_r    
\def\perc_scale#1#2{
  \def\after_setbox{%
    \hbox\bgroup
    \dim_tmp\wd\box_tmp \divide\dim_tmp100 \wd\box_tmp#1\dim_tmp
    \dim_tmp\ht\box_tmp \divide\dim_tmp100 \ht\box_tmp#2\dim_tmp
    \dim_tmp\dp\box_tmp \divide\dim_tmp100 \dp\box_tmp#2\dim_tmp
    \box\box_tmp 
\egroup}%
    \afterassignment\jump_setbox\setbox\box_tmp =
}%
{\catcode`\p12 \catcode`\t12 \gdef\PT_{pt}}
\def\hull_num{\expandafter\hull_num_}
\expandafter\def\expandafter\hull_num_\expandafter#\expandafter1\PT_{#1}
\def\find_scale#1#2{
  \def\after_setbox{%
    \resize\dim_tmp{100pt}{#1}{#2\box_tmp}%
    \xdef\lastscale{\hull_num\the\dim_tmp}\extra_complete}%
  \afterassignment\jump_setbox\setbox\box_tmp =
}
\def\scaleto#1#2#3#4{
  \def\extra_complete{\perc_scale{#3}{#4}\hbox{\box\box_tmp}}%
  \find_scale{#1}#2}
\let\xyscale\perc_scale
\def\zscale#1{\xyscale{#1}{#1}}
\def\yxscale#1#2{\xyscale{#2}{#1}}
\def\xscale#1{\xyscale{#1}{100}}
\def\yscale#1{\xyscale{100}{#1}}
\def\xyscaleto#1{\scaleto{#1}\wd\lastscale\lastscale}
\def\yxscaleto#1{\scaleto{#1}\ht\lastscale\lastscale}
\def\xscaleto#1{\scaleto{#1}\wd\lastscale{100}}
\def\yscaleto#1{\scaleto{#1}\ht{100}\lastscale}
\def\slant#1{
  \hbox\bgroup
  \def\after_setbox{%
    \box\box_tmp 
\egroup}%
  \afterassignment\jump_setbox\setbox\box_tmp =
}%
\def\rotate#1{
  \hbox\bgroup
  \def\after_setbox{%
    \setbox\box_tmp\hbox{\box\box_tmp}
    \wd\box_tmp 0pt \ht\box_tmp 0pt \dp\box_tmp 0pt
    \box\box_tmp
    \egroup}%
  \afterassignment\jump_setbox\setbox\box_tmp =
}%
\newdimen\box_tmp_dim_a
\newdimen\box_tmp_dim_b
\newdimen\box_tmp_dim_c
\def\plus_{+}
\def\minus_{-}
\def\revolvedir#1{
  \hbox\bgroup
   \def\param_{#1}%
   \ifx\param_\plus_ \else \ifx\param_\minus_
   \else
     \errhelp{I would rather suggest to stop immediately.}%
     \errmessage{Argument to \noexpand\revolvedir should be either + or -}%
   \fi\fi
  \def\after_setbox{%
    \box_tmp_dim_a\wd\box_tmp
    \setbox\box_tmp\hbox{%
     \ifx\param_\plus_\kern-\box_tmp_dim_a\fi
     \box\box_tmp
     \ifx\param_\plus_\kern\box_tmp_dim_a\fi}%
    \box_tmp_dim_a\ht\box_tmp \advance\box_tmp_dim_a\dp\box_tmp
    \box_tmp_dim_b\ht\box_tmp \box_tmp_dim_c\dp\box_tmp
    \dp\box_tmp0pt \ht\box_tmp\wd\box_tmp \wd\box_tmp\box_tmp_dim_a
    \kern \ifx\param_\plus_ \box_tmp_dim_c \else \box_tmp_dim_b \fi
    \box\box_tmp
    \kern -\ifx\param_\plus_ \box_tmp_dim_c \else \box_tmp_dim_b \fi
    \egroup}%
  \afterassignment\jump_setbox\setbox\box_tmp =
}%
\def\revolve{\revolvedir-}
\def\xflip{
  \hbox\bgroup
  \def\after_setbox{%
    \box_tmp_dim_a.5\wd\box_tmp
   \setbox\box_tmp
     \hbox{\kern-\box_tmp_dim_a \box\box_tmp \kern\box_tmp_dim_a}%
   \kern\box_tmp_dim_a
    \box\box_tmp
    \kern-\box_tmp_dim_a
    \egroup}%
  \afterassignment\jump_setbox\setbox\box_tmp =
}%
\def\yflip{
  \hbox\bgroup
  \def\after_setbox{%
    \box_tmp_dim_a\ht\box_tmp \box_tmp_dim_b\dp\box_tmp
    \box_tmp_dim_c\box_tmp_dim_a \advance\box_tmp_dim_c\box_tmp_dim_b
    \box_tmp_dim_c.5\box_tmp_dim_c
   \setbox\box_tmp\hbox{\vbox{%
     \kern\box_tmp_dim_c\box\box_tmp\kern-\box_tmp_dim_c}}%
   \advance\box_tmp_dim_c-\box_tmp_dim_b
   \setbox\box_tmp\hbox{%
     \lower\box_tmp_dim_c\box\box_tmp
}%
    \ht\box_tmp\box_tmp_dim_a \dp\box_tmp\box_tmp_dim_b
    \box\box_tmp
    \egroup}%
  \afterassignment\jump_setbox\setbox\box_tmp =
}%
\catcode`\_\undtranscode

\special{ps:}

\newcommand {\eqn}[1]{\begin{equation}#1\end{equation}}
\newcommand {\eqna}[1]{\begin{eqnarray}#1\end{eqnarray}}
\newcommand{\equln}[2]{\begin{equation} {#1} \label{#2} \end{equation}}
\newcommand{\eqnl}[2]{\begin{equation} {#1} \label{#2} \end{equation}}
\newcommand {\la}{\longrightarrow}
{\bf Summary}
    The discussion of enhancement of superconductivity is presented in
the framework of the higher-grade hybrid model of layered
superconductors. The enhancement is considered from the
following two points of view: (i) as a result of
long-range couplings with respects to the short distance ones, and
(ii) the enhancement of 3D superconductivity with respect to the 2D.
 Two important cases are distinguished: homogeneous bulk
supercoducting materials and superconducting structures composed of a
finite number of layers.

\section{Introduction}
    The higher grade hybrid model (HM) of layered superconductors
has been proposed and studied in a series of papers 
\cite{Sztyren:2002}--\cite{Sztyren 2008/2}. It turned out that the
model is also able to describe some phenomena which are beyond the
capabilities offered by simpler phenomenologies such as the 3D anisotropic
Ginzburg-Landau or Lawrence-Doniach theory
\cite{Lawrence+Doniach:71}--\cite{Krasnov:2001}. As an important example
one can mention the phenomenon of enhancement of superconductivity by
interlayer couplings. As a result the layered superconducting structure
can exhibit better superconducting properties than the isolated layers.\\
\indent The present paper is devoted to a more systematic discussion of
such enhancement effects. We shall be concerned with two important
cases: homogeneous bulk supercoducting materials and superconducting
structures composed of a finite number of layers. For the sake of
simplicity we restrict ourselves to situations with no external
magnetic fields.

   In the simplest case the isolated atomic planes are all described
by the model 2D GL with the same parameters $\alpha_0$ and $\beta$.
The parameter $\alpha_0$ depends on the temperature. 
If $\alpha_0>0$, then the 2D state of such a plane is N (normal),
in the oposite case it is the state S (superconducting). We
interpret the parameter $\alpha_0$ as a mesure of empiric
temperature, and introduce the notation $\tau=\tau_0+\alpha_0$.
In consequence, for an isolated plane we have state N for
$\tau>\tau_0$ and state S for $\tau<\tau_0$. 

   For a given material the highest temperature at which the normal
state becomes unstable determines the onset of superconductivity.
Instability of the normal state implies the stability of a
superconducting state. The presence of interactions between planes
modifies the above defined empiric temperature, which becomes a
function of the coupling parameters. In the HM model there are two
classes of long-range coupling constants: Josephson parameters
\(\gamma_q\) and proximity efect parameters \(\zeta_q\), with
$q\epsilon \{1, 2, ..., K\}$. The grade \(K\), expressed
by an arbitrary (but specified for any particular case) integer,
defines the admitted
range of J-links in terms of interplanar gaps.

\section{Infinite medium}
  According to \cite{Sztyren:2003}, in the framework of HM
the layered superconductor is considered
as a one-dimensional stack of 2D layers (eg. atomic
planes) described by 2D Ginzburg-Landau theory with parameters $\alpha_0$
and $\beta$, and interlayer Josephson-type bonds (called J-links)
between them. These couplings represent Josephson's interactions as
well as proximity effect.

   We denote by \(\psi_n\) the order
parameter associated to the layer indexed by the number \(n\). Its
complex conjugate (c.c.) is denoted by \(\bar{\psi}_n\).

    The field equations then have the form
\begin{equation}
  -\frac{\hbar^2}{2m_{ab}}{\bf \nabla}^2\psi_n+\tilde{\alpha}\psi_n+
     \beta|\psi_n|^2\psi_n
    -\frac{1}{2}\sum_q
    \gamma_q
    (\psi_{n+q}+
    \psi_{n-q})=0,
\label{eqpsi}
\end{equation}
where the GL temperature parameter \(\alpha_0\) has been replaced
by $\tilde{\alpha}$, which takes into account the influence of
proximity effect couplings:
\begin{equation}
\tilde{\alpha}=\alpha_0+\frac{1}{2}
      \sum_q\zeta_q.\label{tilal}
\end{equation}
It is seen that $\tilde{\alpha}$ depends on \(n\) for planes near
the boundary (if the number of layers is finite or semi-infinite). 
    We shall pay our attention to the case $K=2$, i.e. the coupling
between nearest and next nearest neighbours. 
Let us consider the ground states of HM. For the plane-uniform
order parameter the field equations have the form
\begin{equation}
\tilde{\alpha}\psi_n +\beta|\psi_n|^2\psi_n-\frac{1}{2}[\gamma_1
(\psi_{n+1}+\psi_{n-1})+\gamma_2(\psi_{n+2}+\psi_{n-2})]=0.
\label{far}
\end{equation}
There exist solutions with constant amplitude and difference of
phase between adjacent atomic planes; they can be obtained by the
ansatz \(\psi_n=C\,e^{in\theta}\), which gives the relation
\eqnl{C^2=-\alpha^*/\beta.
}{uni}
The condition of vanishing Josephson current reads
\eqnl{\gamma_1\sin\theta+2\gamma_2 \sin 2\theta=0.
}{jos}
Solving (\ref{jos}) with respect to $\theta$ we obtain 3 classes of 
ground states, fulfilling (\ref{uni}) with $\alpha^*$ expressed
by coupling parameters:\\
\(\begin{array}{lll}
\bigcirc\hskip-1.2em=\ \ {\rm uniform :} & \ \psi_n=C, & \ \alpha^*=\alpha_0 + \zeta_1 + \zeta_2
-\gamma_1 - \gamma_2,\\
\bigcirc\hskip-1.2em\pm\ \ {\rm alternating:} & \ \psi_n=(-1)^n C, & \ \alpha^*=\alpha_0 + \zeta_1 + \zeta_2
+\gamma_1 - \gamma_2,\\
\bigcirc\hskip-1.2em\approx\ \ {\rm phase\ modulated:} & \ \psi_n=Ce^{\pm in\theta}, &\ \ 
  \theta={\rm \arccos}(-\frac{\gamma_1}{4\gamma_2}).
\end{array}\)\label{arccos}\\\\
The phase modulated solutions exist if the parameters
\(\gamma_1\) and \(\gamma_2\) satisfy the relation
\(|\gamma_1|\,{\leq}4|\,\gamma_2|\).
Then \(\alpha^*\) is expressed by the coupling
constants according to the formula
\begin{equation}
\alpha^*=\alpha_0+\zeta_1+\zeta_2+\gamma_2(1+\frac{\gamma_1^2}
  {8\gamma_2^2}).\label{al3}
\end{equation}
The special symbols in definitions of classes of solutions refer
to the regions at the plane ($\gamma_1,\ \gamma_2$) in which the solutions
are stable:
\eqnl{\bigcirc\hskip-1.2em=\ :\ \gamma_1 >0,\ \gamma_1+4\gamma_2>0,
}{unif}
\eqnl{\bigcirc\hskip-1em\pm\ :\ \gamma_1 <0,\ \gamma_1-4\gamma_2<0,
}{alter}
\eqnl{\bigcirc\hskip-1.2em\approx\ :\ \gamma_2 <0,\ 4\gamma_2<\gamma_1<-4\gamma_2.
}{pham}
The same symbols will be used to mark the corresponding sectors in
the plot below,\\
\vskip 0pt
\noindent \hfil\hbox{\epsffile{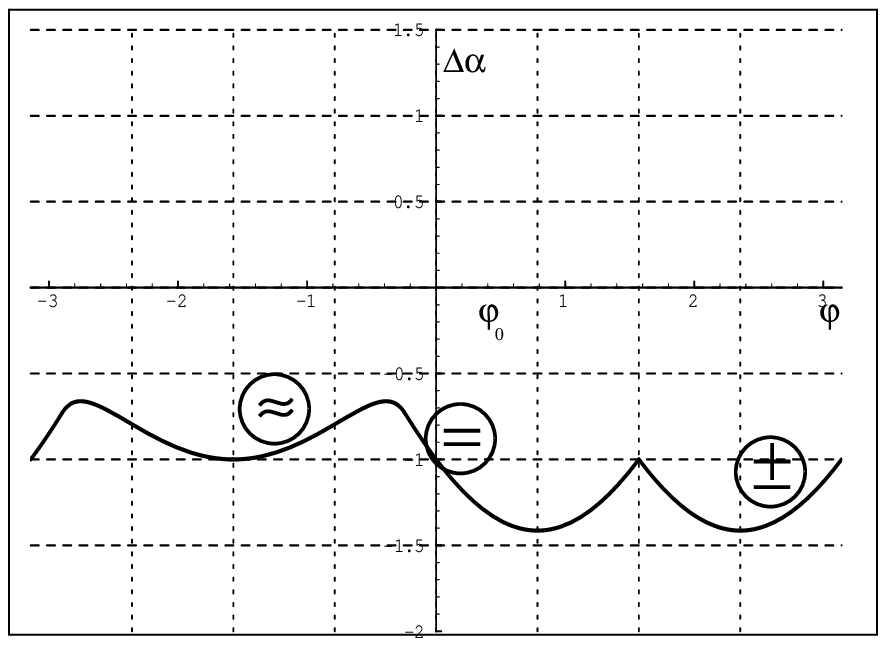}}\hfil\\
\vskip -0.5\baselineskip
\noindent
which illustrates the enhancement of superconductivity. As a measure
of enhancement
we have used the quantity \(\Delta\alpha=\alpha^*-\tilde{\alpha}\)
considered a
function of the coupling angle \(\varphi\) defined as the polar angle
in the plane ($\gamma_1,\ \gamma_2$). The second polar co-ordinate
will be denoted by $\gamma$: 
$\gamma_1=\gamma\cos\varphi,\ \gamma_2=\gamma\sin\varphi$. The numerical 
values have been computed for $\gamma=1$.
The enhancement reaches maximum when \(\Delta\alpha\) has minimum.
We shall denote the minimum value of \(\Delta\alpha\) by $M$.
For the individual classes of solutions we have
\eqnl{\bigcirc\hskip-1.2em=\ :\ \ \Delta\alpha=\sqrt{2}\gamma\sin(\varphi-\frac{3}{4}\pi), 
 \ -\varphi_0\,{\leq}\varphi\,{\leq}\pi-\varphi_0.,\ \ M=
 \Delta\alpha(\varphi/4)=-\sqrt{2}\gamma,
}{eq}
\eqnl{\bigcirc\hskip-1em\pm\ :\ \ \Delta\alpha=\sqrt{2}\gamma\sin(\varphi+
\frac{3}{4}\pi),
 \ \ \varphi_0\,{\leq}\varphi\,{\leq}\,\pi+\varphi_0,\ \ M=
 \Delta\alpha(\frac{3}{4}\pi)= -\sqrt{2}\gamma,
}{pm}
\eqnl{\bigcirc\hskip-1.2em\approx\ :\ \ \Delta\alpha=\gamma(1+\frac{1}{8}
{\rm ctg}^2\varphi)\sin\varphi,
 \ \ -\pi+\varphi_0\,{\leq}\,\varphi \,{\leq}\,-\varphi_0, M=
 \Delta\alpha(-\frac{{\pi}}{2})=-\gamma.
}{app}
In the above formulae the notation
 $\varphi_0=\arctan(\frac14)$ have been used.\\
It follows from the formulae defining the solutions, that for 
appropriate values of coupling parameters one can obtain
negative value of \(\alpha^*\) with positive $\alpha_0$.
Even more,
\(\alpha^*\) can be negative while $\tilde{\alpha}$ remains
positive. For such materials the out-of-plane
superconductivity appears in spite of the fact that all the layers
remain in overcritical in-plane states.

\section{Finite structures}
   According to \cite{Sztyren 2008/2}, one considers
the structure of
3 GL layers coupled\\
by J-links of first and second grade. The order
parameters will be
denoted by \(\psi_{-1},\ \psi_0\) and \(\psi_1\). Their real and
imaginary parts will be denoted by $a$ and $b$ with appropriate
indices.
The field equations have the form
\equln{(\tilde{\alpha}_1+\beta|\psi_1|^2)\psi_1-\frac{1}{2}
 [\gamma_1\psi_0+\gamma_2\psi_{-1}]=0,}
{c3_1}
\equln{(\tilde{\alpha}_0+\beta|\psi_0|^2)\psi_0-\frac{1}{2}
 \gamma_1(\psi_1+\psi_{-1})=0,}
{c3_2}
\equln{(\tilde{\alpha}_{-1}+\beta|\psi_{-1}|^2)\psi_{-1}-\frac{1}{2}
 [\gamma_1\psi_0+\gamma_2\psi_1]=0,}
{c3_3}
where 
\equln{\tilde{\alpha}_0=\alpha_0+\zeta_1
.}{alf0}
and
\equln{\tilde{\alpha}_1=\tilde{\alpha}_{-1}=\alpha_0+\frac{1}{2}(\zeta_1+\zeta_2)
              =\tilde{\alpha}_0-\frac12\delta
,}{alf1}
with
\equln{\delta=\zeta_1-\zeta_2.
}{delta}
From the form of equations it follows that any solution is 
gauge-equivalent to a triplet $\psi_{-1},\ \psi_0,\ \psi_1$, for which
\eqnl{b_0=0,\ \ b_1=-b_{-1},\ \ b_1[\gamma_1a_0+\gamma_2(a_1+a_{-1})]=0.
}{prop1}
For any real solution either $a_1=a_{-1}$, or
\eqnl{\alpha^*_1+\beta(a_{1}^2+a_1a_{-1}+a^2_{-1})=0,
}{prop2}
where
\eqnl{\alpha^*_1=\tilde{\alpha}_0+\frac{1}{2}(\gamma_2-\delta).
}{pro2a}

   The set of nontrivial solutions to the 
system (\ref{c3_1}-\ref{c3_3}) is
partitioned into four disjoint classes, two of which are specially
interesting from the point of view of the onset and enhancement of
superconductivity. Namely the class\\
\indent 
(A) containing solutions fulfilling conditions 
\eqnl{b_1=0,\ \ a_0=0,\ \ a_1=-a_{-1}\,{\neq}\,0,\\ 
}{clA}
and the class\\
\indent 
(C) with solutions fulfilling conditions 
\eqnl{b_1=0,\ \ a_0\,{\neq}\,0,\ \ a_1=a_{-1}.
}{clC}
The solutions of the class (A) may be explicitly written as
\equln{a_1^2=-\frac{\alpha^*_1}{\beta}
,}{a1}
so that (taking into account the positiveness of \(\beta\))
the necessary and sufficient condition for existence of mode A
is 
\equln{2\tilde{\alpha}_0+\gamma_2-\delta<0
.}{lin}
For the class (C) the system of equations (\ref{c3_1}-\ref{c3_3})
may be transformed into the form
\equln{a_0=\frac 2{\gamma_1}(\alpha^*_1-\gamma_2+\beta a_1^2)a_1
,}{c1}
\equln{a_1=\frac 1{\gamma_1}(\tilde{\alpha}_0+\beta a_0^2)a_0
,}{c2}
where $\alpha^*_1$ is given by the eqn. (\ref{pro2a}). 
Introducing
\equln{x=\tilde{\alpha}_0+\beta a_0^2
,}{anx}
one can express the necessary conditions for the existence of mode C:
\equln{\beta a_0^2=x-\tilde{\alpha}_0>0
}{betaa0}
and
\equln{\beta a_1^2=\gamma_1^2\frac{1}{x}-\alpha_1^*-\gamma_2>0
.}{betaa1}
Further
\equln{\frac{x^3}{\gamma_1^2}=
\frac{\gamma_1^2-2\tilde{\alpha}_0+
(\delta+\gamma_2)}{2(x-\tilde{\alpha}_0)}
.}{x22}
Hence, geometrically, the solutions from class (C) are determined
by the points of intersection of two curves:
\equln{y=\frac{2}{\gamma_1^2}x^3
,}{y1}
and
\equln{y=\frac{\gamma_1^2-(2\tilde{\alpha}_0
-\delta-\gamma_2)x}{x-\tilde{\alpha}_0}
.}{y2}
From the equations (\ref{c1}-\ref{c2}) and (\ref{anx}) it follows
that for $x=\tilde{\alpha}_0$ we have the zero-solution,
hence the transition to the normal state. The $\tilde{\alpha}_0$
fulfills the equation
\equln{(2\tilde{\alpha}_0-\delta-\gamma_2)\tilde{\alpha}_0
-\gamma_1^2=0
,}{Ns}
which always has two real roots. We shall denote them by
\equln{\alpha_{01}=\frac 14(\delta+\gamma_2-\sqrt{(\delta+\gamma_2)^2+
8\gamma_1^2})
,}{al1}
and
\eqnl{\alpha_{02}=\frac 14(\delta+\gamma_2+\sqrt{(\delta+\gamma_2)^2+
8\gamma_1^2})
.}{al2}
The class (C) is empty if 
\eqnl{\tilde{\alpha}_0>\alpha_{02}
.}{nor1}
\noindent
When $\tilde{\alpha}_0<\alpha_{01}$, there exist two solutions.\\
When $\alpha_{01}<\tilde{\alpha}_0<\frac 12(\delta+\gamma_2)$
or $\frac 12(\delta+\gamma_2)<\tilde{\alpha}_0<\alpha_{02}$,
there is one solution.\\\\

   The temperature relation (\ref{nor1}) is one of the set of conditions
for stability of the normal state. From the analysis of the second
variation of energy of the normal state it follows that the remaining
two conditions have the form $\tilde{\alpha}_0>0$ and 
$\tilde{\alpha}_0>\frac 12(\delta-|\gamma_2|)$.

    When the normal mode becomes unstable, either
mode A (onset A) or mode C (onset C) can apear, dependig on the position
in the material plane ($\gamma_2,\delta$). The plane is divided into
two regions of different onset by the curve
\eqnl{\delta=\gamma_2-\frac{\gamma_1^2}{\gamma_2},\ \ \gamma_2<0
.}{limi}
The following combinations of the material parameters define
two empiric temperatures
\equln{\tau_{A}=\tau_0-\zeta_1+\frac 12(\delta-\gamma_2)
,}{tAN}
and
\equln{\tau_{C}=\tau_0-\zeta_1+\alpha_{02}
,}{tCN}
where $\alpha_{02}$ is given by the eqn. (\ref{al2}).
As long as $\tau>$max$(\tau_A, \tau_C)$, the normal state is stable.
Below this limit a stable superconducting mode apeares.
If  $\tau_{A}>\tau_{C}$, the mode is A; in the
opposite case -- it is the mode C. The quantity $\tau_0$ describes
the onset for isolated GL planes: if $\tau>\tau_0$ then the planes are
in normal state, if $\tau<\tau_0$ then the state is superconducting.\\

   Let us now examine the increment of the onset temperatures
$\tau_A$ and $\tau_C$ due to the long distance couplings. Denoting
the onset temperatures for the first
grade material by
$\tau_{A0}$ and $\tau_{C0}$ , one can calculate
\equln{\tau_{A}-\tau_{A0}=-\frac 12(\zeta_2+\gamma_2)
,}{t0N}
and
\equln{\tau_{C}-\tau_{C0}=-\frac 14[
\sqrt{(\zeta_1-(\zeta_2-\gamma_2))^2+8\gamma_1^2}-
\sqrt{\zeta_1^2+8\gamma_1^2}+(\zeta_2-\gamma_2)]
.}{t0C}
Hence the mode A supercoductivity is enhanced provided that 
$\zeta_2+\gamma_2<0$. On the other hand, the mode C superconductivity
is enhanced provided that $\zeta_2-\gamma_2>0$.

   It follows from the formulae (\ref{tAN}) and (\ref{tCN})
that for appropriate values of parameters $\zeta_1$, $\zeta_2$ and
$\gamma_2$, the onset
temperatures $\tau_A$ and/or $\tau_C$ may become greater than
the 2D critical
temperature $\tau_0$. For such a material the structure may be
supercoducting even
if it is composed of planes which after separation are
overcritical.

\newpage
\thebibliography{12}
\bibliography{}
\end{document}